\documentclass[journal, a4paper]{IEEEtran}

\usepackage{cite}      

\usepackage{graphicx}   

\usepackage{amsmath,amssymb}   

\usepackage{tikz,pgfplots,pgfplotstable}
\usetikzlibrary{calc, intersections}
\usetikzlibrary{shapes.geometric}
\usetikzlibrary{positioning}

\usepackage{comment}

\newlength\figH
\newlength\figW
\pgfplotsset{every tick label/.append style={font=\footnotesize}, every axis label/.append style={font=\footnotesize}, legend style = {font=\footnotesize}}

\usepackage[normalem]{ulem}

\newcommand{\new}[1]{#1}
\newcommand{\newUn}[1]{#1}
\pgfplotsset{compat=1.13}
\usetikzlibrary{external}
\tikzexternalenable

\begin{document}

\title{Experimental Comparison of Probabilistic Shaping Methods for Unrepeated Fiber Transmission}
\author{Julian~Renner, Tobias~Fehenberger,~\IEEEmembership{Student Member,~IEEE}, Metodi~P.~Yankov,~\IEEEmembership{Member,~IEEE}, Francesco~Da~Ros,~\IEEEmembership{Member,~IEEE}, S{\o}ren Forchhammer,~\IEEEmembership{Member,~IEEE}, Georg~B{\"o}cherer,~\IEEEmembership{Member,~IEEE}, and~Norbert~Hanik,~\IEEEmembership{Senior~Member,~IEEE}
\thanks{This work was supported by the DNRF Research Center of Excellence, SPOC (ref. DNRF123), by the German Federal Ministry of Education and Research in the framework of an Alexander von Humboldt Professorship, and by a fellowship within the FITweltweit program of the German Academic Exchange Service (DAAD).}
\thanks{Julian Renner was with the Institute for Communications Engineering, Technical University of Munich (TUM), 80333 Munich, Germany and is now with the Signal Processing Systems Group, Eindhoven University of Technology (TU/e), 5612 AZ Eindhoven, Netherlands (Email: j.w.renner@tue.nl).}
\thanks{Tobias Fehenberger, Georg~B{\"o}cherer  and Norbert Hanik are with the Institute for Communications Engineering, Technical University of Munich (TUM), 80333 Munich, Germany (\mbox{Emails:}~tobias.fehenberger@tum.de, georg.boecherer@tum.de, norbert.hanik@tum.de).}
\thanks{Metodi~P.~Yankov,  Francesco~Da~Ros and S{\o}ren Forchhammer are with the Department of Photonics Engineering, Technical University of Denmark, 2800 Kgs. Lyngby, Denmark (\mbox{Emails:}~meya@fotonik.dtu.dk, fdro@fotonik.dtu.dk and  sofo@fotonik.dtu.dk).}
} 

\maketitle

\begin{abstract}
This paper studies the impact of probabilistic shaping on effective signal-to-noise ratios (SNRs) and achievable information rates (AIRs) in a back-to-back configuration and in unrepeated nonlinear fiber transmissions. For back-to-back, various shaped quadrature amplitude modulation (QAM) distributions are found to have the same implementation penalty as uniform input. By demonstrating in transmission experiments that shaped QAM input leads to lower effective SNR than uniform input at a fixed average launch power, we experimentally confirm that shaping enhances the fiber nonlinearities. However, shaping is ultimately found to increase the AIR, which is the most relevant figure of merit as it is directly related to spectral efficiency. In a detailed study of these shaping gains for the nonlinear fiber channel, four strategies for optimizing QAM input distributions are evaluated and experimentally compared in wavelength division multiplexing (WDM) systems. The first shaping scheme generates a Maxwell-Boltzmann (MB) distribution based on a linear additive white Gaussian noise channel. The second strategy uses the Blahut-Arimoto algorithm to optimize an unconstrained QAM distribution for a split-step Fourier method based channel model. In the third and fourth approach, MB-shaped QAM and unconstrained QAM are optimized via the enhanced Gaussian noise (EGN) model. \newUn{Although the absolute shaping gains are found to be relatively small, the relative improvements by EGN-optimized unconstrained distributions over linear AWGN optimized MB distributions are up to 59\%. This general behavior is observed in 9-channel and fully loaded WDM experiments.}

\end{abstract}


\section{Introduction}

\IEEEPARstart{I}{n} optical fiber communications, nonlinear interference (NLI) limits the effective signal-to-noise ratio (SNR) at the receiver, which in turn confines the maximum spectral efficiency (SE) achievable in current transmission systems. In order to facilitate the continuously growing data rates of fiber-optical links \cite{Bayvel2016PhilTransRSocA_MaximizingCapacityReview}, techniques must be sought that offer increased SEs. As both electrical and optical power can be limited in long-haul fiber links \cite{Frisch2013ProcSuboptic_ElectricalPower,Maher2016ECOC_Submarine} and the electronic components in the transceiver restrict the maximum effective SNR \cite{Maher2015ECOC_GMICurves} even in a back-to-back configuration, the power efficiency of any throughput-increasing method is of pivotal importance. Constellation shaping is a technique that fulfills this requirement, and particularly probabilistic shaping has recently attracted a lot of research attention in fiber optics. 

For the linear additive white Gaussian noise (AWGN) channel, it is well-known that probabilistic shaping achieves large gains up to $1.53$~dB effective SNR \cite{Forney1984JSEL_Shaping,Fischer2005Book_Shaping}, and even larger improvements are potentially for the nonlinear fiber channel \cite{Dar2014}. For linear AWGN channels, input constellations that follow Maxwell-Boltzmann (MB) distributions are often used as they allow to operate closely to the AWGN capacity \cite{KschischangPasupathy1993,bocherer2015,Wu2012_optMB}. \newUn{This family of distributions is also frequently applied in fiber transmissions, see, e.g.,}\new{\cite{SmithKschischang2012,Beygi2014JLT_CodedModulation,Fehenberger2015,Fehenberger2016PTL_MismatchedShaping,Buchali2016,Cho2016ECOC_ShapingNonlinearTolerance}}\newUn{, thus enabling large performance improvements in both reach and SE. The impact of MB distributions on the fiber NLI and thus, on the effective SNR, has, for example, been studied and compared to other distributions in}\new{\cite{Cho2016ECOC_ShapingNonlinearTolerance,PanKschischang2016_ShapingNonlinearChannel,Fehenberger2016, Yankov16_exp}}\newUn{, and only relatively minor improvements are found by non-MB distributions for the considered multi-span links with lumped amplification. The non-MB distributions are found via fourth-moment optimization}\new{\cite{Cho2016ECOC_ShapingNonlinearTolerance}}\newUn{, enhanced Gaussian noise (EGN) models}\new{\cite{PanKschischang2016_ShapingNonlinearChannel, Fehenberger2016}}\newUn{, or the Blahut-Arimoto (BA) algorithm with a symmetrized split-step Fourier method (SSFM) based channel model}\new{\cite{Yankov2014PTL_ProbabilisticShaping, Yankov16_exp}}\newUn{. For short-reach applications, a shell mapping algorithm is used in}\new{\cite{Geller2016}}\newUn{ to jointly shape symbols that are transmitted over consecutive time slots in the same wavelength division multiplexing (WDM) channel. This concept of temporal shaping is also investigated in}\new{\cite{Yankov2017_TempShaping}}\newUn{ with finite state machines. We restrict the presented analysis to probabilistic shaping on a symbol-by-symbol basis to keep the receiver complexity low.}

For multi-span long-haul transmission systems, it has been shown via EGN models and in simulations that shaping with MB distributions decreases the effective SNR but increases the achievable information rate (AIR) \cite{Fehenberger2016}. The reason for the SNR reduction is that, at a fixed average launch power, shaping enhances the nonlinear fiber effects. The magnitude of this effect, however, was found to be relatively small for dispersion-unmanaged long-haul transmission \cite[Sec.~IV-E]{Fehenberger2016}. Closely related to this finding, it was further shown in simulations that for such links the AIR curves of all optimized distributions are expected to be on top of each other \cite[Sec.~IV-F]{Fehenberger2016}.
In conclusion, an overall performance improvement in terms of AIRs is observed, which are directly related to the maximum SE of a particular coded modulation system and thus the more significant figures of merit than effective SNR \cite{Alvarado2015JLT_BitWise,Cartledge2017}.

\newUn{For unrepeated optical transmission systems, no in-line active elements are required, which means that they provide a cost-efficient solution for short links and are thus crucial for the Internet connectivity of isolated communities and island-to-island or island-to-continent connections}\new{\cite{Januario_intro},\cite{Galdino_intro}}\newUn{. As a result of the growing capacity demands, increasing the throughput for such systems is of a high relevance. This paper experimentally studies the impact of various probabilistically shaped input distributions on SNR and AIR. Since the impact of the input distribution on the NLI at the optimal launch power is small for multi-span long-haul transmissions}\new{\cite{Fehenberger2016}}\newUn{, the focus of this paper is on unrepeated transmission systems for which the optimal launch power is generally much higher, resulting in a stronger relative contribution of the NLI to the overall transmission impairments.} 

The main contributions of this paper are as follows. Firstly, it is demonstrated in back-to-back experiments that there is no additional implementation penalty from using probabilistically shaped input constellations. Secondly, unrepeated transmission experiments with a variety of optimized input distributions are conducted and confirm that shaping decreases SNR but increases AIR. Thirdly, it is shown that the considered shaped inputs give improvements for a $9$ WDM channels transmission as well as for a fully loaded $80$ WDM channels transmission.

\section{Probabilistic Shaping for Fiber Channels}\label{sec:shaped_dists}
\subsection{Optimization Strategies}
\begin{table}
        \caption{System parameters for PMF optimization}
        \begin{center}
                \begin{tabular}{ r | l }
                        Nonlinear coefficient $\gamma$ & $1.3$ $1$/(W$\cdot$km) \\
                        Chromatic dispersion $D(\lambda)$  & $17$ ps/(nm$\cdot$km) \\
                        Attenuation $\alpha$ & $0.2$ dB/km \\
                        Amplification & EDFA \\
                        EDFA noise figure $n_{\text{F}}$ & $5$ dB \\
                        Laser wavelength $\lambda_{0}$ & $1.55$ $\mu$m \\
                        Number of spans $N_{\text{span}}$ & $1$ \\
                        Span length $l_{\text{span}}$ & $200$ km \\
                        Step size $h$ & $0.1$ km \\
                        Oversampling factor $f_{\text{os}}$  & $32$ \\
                        Number of symbols $N_{\text{Sym}}$ &  $1\cdot 10^5$ per WDM channel \\
                \end{tabular}
        \end{center}
        \label{tab:system_parameters_sim_for_exp1_64}
\end{table}
Four different methods to find optimized shaped quadrature amplitude modulation (QAM) input distributions for the nonlinear fiber channel are compared. The first method is described in \cite{Fehenberger2015} and is herein referred to as linear Maxwell-Boltzmann (Lin-MB). In this shaping method, a uniform distribution is transmitted over the fiber system and the effective SNR is estimated at the receiver. The estimated effective SNR is used to determine an optimized MB distribution under the assumption of a linear AWGN channel \cite{bocherer2015}. The family of MB distributions correspond to sampled Gaussian densities \cite{KschischangPasupathy1993}, whose probability mass function (PMF) is given by
\begin{align}
P^*_X(x_i) = \frac{1}{\sum_{j=1}^{M} \exp(-\nu |x_{j}|^2)}\exp(-\nu |x_{i}|^2).
\end{align}
The random variable $X$ represents the channel input with realizations $x_1,\hdots,x_M$, $M$ is the size of the input alphabet and $\nu$ denotes a scaling factor. Since the distribution is optimized under the assumption that a change of the input constellation does not effect the SNR, the SNR does not have to be updated in each optimization iteration. Furthermore, an MB distribution is fully described by $\nu$ and thus requires a single optimization parameter. These facts make the Lin-MB very efficient for optimization \cite[Sec.~III-C]{bocherer2015}.

The second method is called SSFM-BA shaping \cite{Yankov16_exp}. The BA algorithm finds the capacity-achieving PMF for a fixed constellation size and the AWGN channel \cite{Varnica2002_CapAWGN}. In this work, the input distribution of a 2D square QAM is optimized by the BA algorithm over an SSFM-based channel model. In contrast to the above MB distributions, the result of the SSFM-BA optimization can be a PMF of any kind, giving a maximum degree of freedom in choosing a suitably shaped distribution. This, however, comes at the expense of huge computational complexity as the BA algorithm is an iterative algorithm, meaning that a full SSFM-based transmission has to be simulated for each iteration.

The third and fourth methods proposed here are referred to as the EGN-MB and EGN-2D, respectively. Both methods use an EGN channel model \cite{Dar:13,Carena2014OptExp_EGNmodel} whose effective SNR after fiber propagation and matched filtering is defined as \cite{Poggiolini2012JLT_GNModel}
\begin{align}
\text{SNR}_{\text{EGN}} = \frac{P_{\text{tx}}}{\sigma_{\text{eff}}^2} = \frac{P_{\text{tx}}}{\sigma_{\text{ASE}}^2 + \sigma_{\text{NLI}}^2 },
\label{eq:effective_SNR}
\end{align}
where $P_{\text{tx}}$ is the launch power, $\sigma_{\text{ASE}}^2$ denotes the noise due to the Erbium-doped fiber amplifiers (EDFAs) and $\sigma_{\text{NLI}}^2$ represents the noise due to NLI. The EGN channel model includes the NLI as additive, isotropic, memoryless, zero-mean Gaussian noise. The term $\sigma_{\text{ASE}}^2$ is computed according to \cite{Essiambre2010}, and the nonlinear noise term of the model can be expressed as \cite{Fehenberger2016}
\begin{align}
\sigma_{\text{NLI}}^2   = P_{\text{tx}}^3\big[\chi_0 + (\hat{\mu}_4-2) \chi_4 + (\hat{\mu}_4-2)^2  \chi_4' +\hat{\mu}_6  \chi_6\big], 
\label{eq:NLi_noise}
\end{align}
where the parameters $\chi_0$, $\chi_4$, $\chi_4'$ and $\chi_6$ are real coefficients that represent the contributions of the fiber nonlinearities and $\hat{\mu}_4 = \mathbb{E}[|X|^4]$ and $\hat{\mu}_6= \mathbb{E}[|X|^6]$ are the fourth and sixth moment of a unit-energy square QAM, respectively.
\begin{table}
        \caption{Summary of the shaping methods}
        \begin{center}
        \setlength{\tabcolsep}{4pt} 
        \renewcommand{\arraystretch}{1.2}
                \begin{tabular}{ r | c | c | c | c }
                        &  Lin-MB & EGN-MB & EGN-2D & SSFM-BA  \\
                        \hline
                        \hline
                        Channel model & AWGN & \multicolumn{2}{c|}{EGN}  & SSFM \\
                        \hline
                        PMF Family  & \multicolumn{2}{c|}{MB} & \multicolumn{2}{c}{any 2D}  \\
                        \hline
                        Degrees of freedom  & \multicolumn{2}{c|}{1 ($\nu$)} &  \multicolumn{2}{c}{$\#$constellation points -1} \\
                        \hline
                        Complexity (++ lowest) & ++ & + & - & - - 
                \end{tabular}
        \end{center}
        \label{tab:shaping-methods}
\end{table}
For the parameters in Table~\ref{tab:system_parameters_sim_for_exp1_64}, four-wave mixing is numerically found to be negligible in comparison to self- and cross-channel interference and is thus not included in \eqref{eq:NLi_noise}. Since $\chi_0$, $\chi_4$, $\chi_4'$ and $\chi_6$ are independent of $P_{\text{tx}}$ and $P_X$, these coefficients have to be computed only once at the beginning of the optimization and are kept constant for each optimization iteration. This means that per optimization iteration, a simple update of $\hat{\mu}_4$ and $\hat{\mu}_6$ is required to determine the SNR, which makes the optimization of $P_X$ based on the EGN model efficient. 
The interior point algorithm \cite{Forsgren_2002} is used to optimize the input distribution such that mutual information is maximized for an AWGN channel whose SNR is determined by \eqref{eq:effective_SNR}. 

Based on the EGN model \eqref{eq:effective_SNR} and \eqref{eq:NLi_noise}, two different families of shaped distributions are investigated. The EGN-MB shaped distribution is constrained to be from the family of MB distributions, whereas the EGN-2D approach allows to create any 2D input whose constellation points are equally spaced, such as multi-ring modulation. The latter approach requires 2D processing in the demapper because the in-phase and quadrature components cannot be separated without loss of information. \newUn{We also investigated a 2D distribution that is the Cartesian product of the 1D distributions that has the smallest Kullback-Leibler divergence to the original EGN-2D. Such a product distribution allows separation of the in-phase and quadrature components in the demapper. Although the decrease in Kullback-Leibler divergence is only 0.04 bit, SSFM simulations (not included here) show a significant shaping gain reduction by approximately 33\% as the new 2D distribution is not directly optimized for the nonlinear fiber channel.} The properties of all four methods are summarized in Table~\ref{tab:shaping-methods}. For the experiments, we choose the more practical approach of optimizing all input distributions at the optimal launch power and keep the distributions constant for each power sweep. This implies that the distributions are suboptimal for nonoptimal power levels. 

\subsection{Figures of Merit for Comparing the Shaping Methods}
The shaping strategies described in the preceding section are evaluated using two different figures of merit. The effective SNR represents the signal-to-noise ratio after the receiver digital signal processing (DSP) chain. We estimate the effective SNR by
\begin{align}
\text{SNR}_{\text{eff}} \approx \frac{\sum_{i=1}^{M} P_{X}(x_i) |\mu_i|^2    }{\sum_{i=1}^{M} P_{X}(x_i) \sigma^2_i  },
\label{eq:SNR_eff}
\end{align}
where $\mu_i$ and $\sigma^2_i$ are empirical mean and variance of the received data symbols that belong to the input data symbol $x_i$ and $P_X(x_i)$ is its probability of occurrence. The $\text{SNR}_{\text{eff}}$ (\ref{eq:SNR_eff}) includes all distortions between channel input and output, i.e., the implementation penalty with DSP imperfections for back-to-back, and additionally all fiber impairments in case of transmission experiments. The focus of this study is on symbolwise AIRs which are estimated with the mismatched decoding technique \cite{Secondini2013} using a memoryless 2D isotropic Gaussian auxiliary channel \cite{Eriksson2016}, i.e.,
\begin{align}
\text{AIR} \approx \frac{1}{N_{\text{Sym}}}  \sum_{k=1}^{N_{\text{Sym}}} \log_{2} \frac{q_{Y|X}(\underline{y}_k|\underline{x}_k) }{q_{Y}(\underline{y}_k)},
\label{eq:AIR_estimation}
\end{align}
where $\underline{x}_k$ and $\underline{y}_k$ denote the $k$\textsuperscript{th} out of $N_{\text{Sym}}$ elements in the sequence of the channel input data symbols and output data symbols, respectively. The Gaussian auxiliary channel with total variance $\sigma_i^2$ is taken as conditionally dependent on the $i$\textsuperscript{th} constellation point as
\begin{align}
q_{Y|X}(y|x_i) = \frac{1}{\pi \sigma_i^2} e^{-\frac{|y-x_i|^2}{\sigma_i^2}}
\end{align}
and 
\begin{align}
q_{Y}(y) = \sum_{i=1}^{M} P_{X}(x_i) q_{Y|X}(y|x_i).
\end{align}
As long as the input distributions are symmetric, we can apply constant composition distribution matching \cite{Schulte2016} in combination with the probabilistic amplitude-shaping scheme \cite{bocherer2015} to construct the input distributions and virtually achieve the shaping gains.


\section{Experimental Setup} \label{section3}
\begin{figure*}
\begin{center}
\includegraphics[width=1\textwidth]{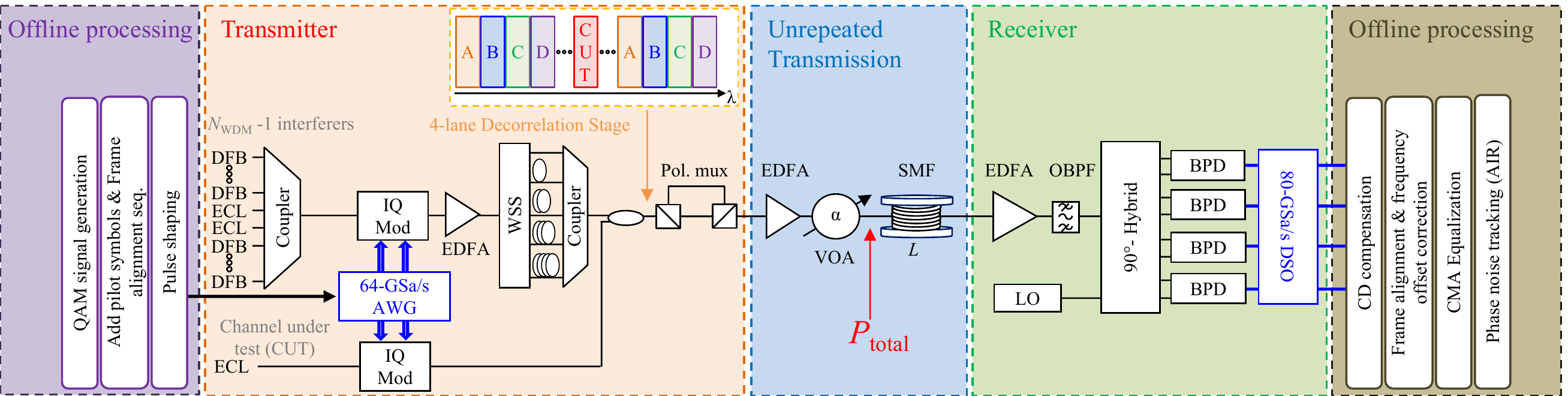}
\caption{Block diagram of the experimental setup. The waveform of one polarization is generated offline and fed to the AWG. The WDM signal is then generated and sent to the unrepeated transmission link. After $80$-GSa/s coherent reception, the received samples are processed offline.}
\label{draw:exp_setup}
\end{center}
\end{figure*}
The experimental setup is shown in Fig.~\ref{draw:exp_setup}. At the transmitter side, QAM data symbols are randomly generated and interleaved with quaternary phase shift keying (QPSK) pilots at a pilot rate of $10\%$. \newUn{Even though the relatively high pilot ratio selected in the experiment results in a loss in net throughput, it was chosen to ensure the stability of the DSP for the duration of the measurements. As demonstrated in}\new{\cite{Yankov2017_ExperimentsPilotbasedEqualization}}\newUn{, the DSP employed in this experiment is independent of the modulation format, and both shaped and uniform constellations exhibit a similar optimal pilot ratio.} A frame alignment sequence is added at the beginning of the sequence and square root raised cosine (RRC) pulse shaping is then applied with a roll-off factor of $0.5$. An arbitrary waveform generator (AWG, $64$-GSa/s and $20$-GHz analog bandwidth) performs the digital-to-analog conversion and drives two IQ modulators used for the central channel (channel under test, CUT) and the $N_{\text{WDM}}-1$ interfering channels, respectively. 

The central channel uses an external cavity laser (ECL, $100$-kHz linewidth), similarly to its two nearest neighbor interferers. The remaining $N_{\text{WDM}}-3$ interferers are based on distributed feedback lasers (DFBs, $1$-MHz linewidth). Two different scenarios are investigated in this work: a $9$-channel WDM system ($N_{\text{WDM}}=9$) based on a $25$-GHz grid and a fully loaded system based on $80$ WDM channels ($N_{\text{WDM}}=80$) on a $50$-GHz grid. The $N_{\text{WDM}}-1$ interferers are amplified in an EDFA and decorrelated in a $4$-lane decorrelation stage based on a wavelength selective switch (WSS). As shown in Fig.~\ref{draw:exp_setup}, the decorrelation scheme follows an ABCD-map ensuring that only every fourth interfering channel ($>100$-GHz apart) contains correlated data. Additionally, the CUT contains a fifth data sequence effectively decorrelating it from all the interferers. After combining the CUT and interferers, a delay-and-add polarization emulator (Pol. mux) is used to generate a dual-polarization signal. 

At the input of the unrepeated transmission link, an EDFA and a variable optical attenuator (VOA) are used to set the total power $P_{\text{total}}$ launched into the standard single-mode fiber (SMF) of length $L$. After transmission, the central channel is amplified in an EDFA, selected by an optical bandpass filter (OBPF) and received using a standard pre-amplified coherent receiver with an additional ECL ($100$-kHz linewidth) acting as a local oscillator (LO). The coherent receiver is followed by a digital storage oscilloscope (DSO, $80$-GSa/s and $33$-GHz analog bandwidth) performing the analog-to-digital conversion. Offline DSP is performed consisting of (in order) low-pass filtering, down sampling, chromatic dispersion (CD) compensation in the frequency domain, frequency offset estimation based on the pilots, time-domain equalization using a pilot-based constant modulus algorithm (CMA) and carrier phase recovery with a trellis-based method \cite{Yankov16_exp,Yankov2017_ExperimentsPilotbasedEqualization}. \newUn{The AIRs are estimated with (\ref{eq:AIR_estimation}). In case of the 9-channel WDM system, the estimation is based on $20$ to $60$ blocks of $12000$ data symbols per channel (i.e, $2.4 \cdot 10^5$ to $7.2 \cdot 10^5$ symbols for the channel under test). For the fully loaded system, $40$ to $120$ blocks of $12000$ data symbols were used, corresponding to a total of $4.8 \cdot 10^5$ to $1.44 \cdot 10^6$ symbols.}\footnote{\newUn{The number of detected blocks is lower bounded by the memory of the AWG, and upper bounded by the length of the useful part of the detected sequence, for which the DSP has converged. In the 20 GBd case, twice as many blocks can fit in the AWG memory due to the smaller oversampling factor.}}

The AIRs of idealized simulations are in general higher than in experiments due to transceiver penalties. The target distances in the experiments were therefore decreased with respect to those of the simulations such that the peak AIR in simulations and the AIR in experiments matched at the optimal launch power. This approach ensures that the signal is shaped for the correct target AIR.

\section{Experimental Results} \label{section4}
\subsection{Optical Back-to-back Results}\label{subsection:A}
\begin{table}
        \caption{Signal parameters of the $9$ WDM channels experiments}
        \begin{center}
                \begin{tabular}{ r | l }
                        Modulation & $64$QAM / $256$QAM \\
                        Symbol rate $R_{X}$ & $10$ GBd \\
                        Pulse shaping & RRC filter \\
                        RRC roll-off $\rho$ & $0.5$ \\
                        WDM channel number $N_{\text{WDM}}$ & $9$ \\
                        WDM channel spacing $f_{\text{WDM}}$ & $25$ GHz \\
                        Polarization & dual \\
                        \newUn{Block length} & \newUn{$13200$ symbols} \\
                        \newUn{Number of blocks} & \newUn{$20$ -- $60$ } \\
                        Pilot rate & $10\%$ \\
                        Channel of interest & Central channel
                \end{tabular}
        \end{center}
        \label{tab:signal_parameters_exp1_64}
\end{table}
\begin{figure}
        \begin{center}
                \includegraphics{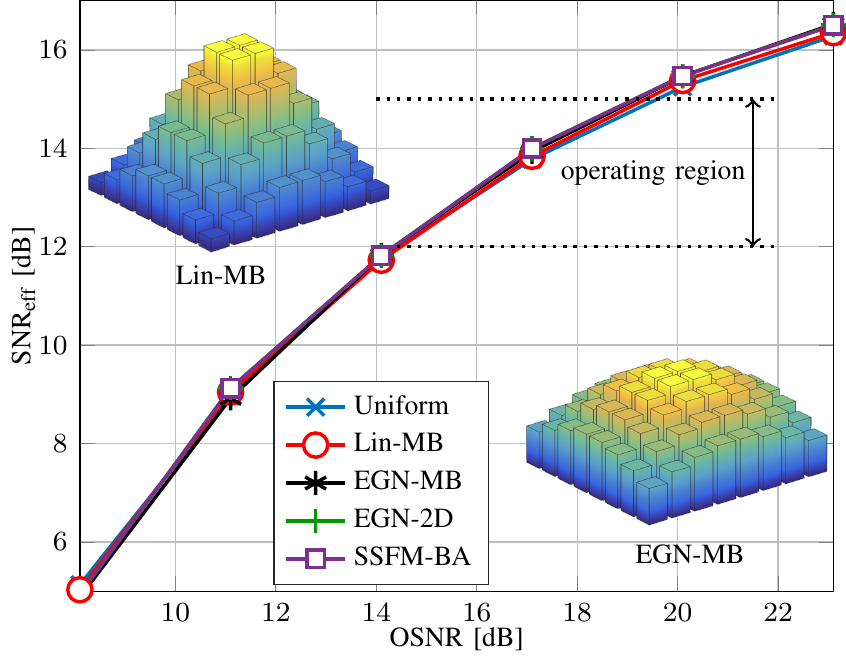}
        \end{center}
        \vspace{-0.5cm}      
        \caption{Estimated received effective SNR as a function of the OSNR in optical back-to-back measurements for $64$QAM. Insets: \newUn{Lin-MB and EGN-MB shaped input distributions.}}
        \label{Exp:b2b_10gbaud_SNR_64}
\end{figure} 
In order to quantify the implementation penalty of the used setup, we conducted $64$QAM optical back-to-back measurements for which the signal parameters are given in Table \ref{tab:signal_parameters_exp1_64}. In Fig.~\ref{Exp:b2b_10gbaud_SNR_64}, the received effective SNR estimated at the end of the DSP chain is shown as a function of the optical signal-to-noise ratio~($\text{OSNR}$) measured over a $0.1$-nm bandwidth. In order to vary the OSNR, a noise loading stage based on an ASE noise source is added in front of the coherent receiver. In an ideal system, the SNR is a linear function of the OSNR (both in dB). In Fig.~\ref{Exp:b2b_10gbaud_SNR_64}, it is shown that for a low OSNR, the received SNR increases linearly with the OSNR in the used setup. However, the slope of the SNR decreases for higher OSNR due to transceiver and DSP imperfections. In our system, the effective SNR was mainly influenced by the finite resolution of the digital-to-analog converter in the transmitter and the finite resolution of the analog-to-digital converter in the receiver. All shaping methods in this paper are found to have the same implementation penalty \newUn{in the operating region}, indicating that with the used components and modulation format independent algorithms, such as the pilot based equalization and synchronization considered here, no penalty from using different shaped input distributions is observed.

%
\subsection{64QAM Transmission with 9 WDM Channels}

\begin{figure*}
\centering
\includegraphics{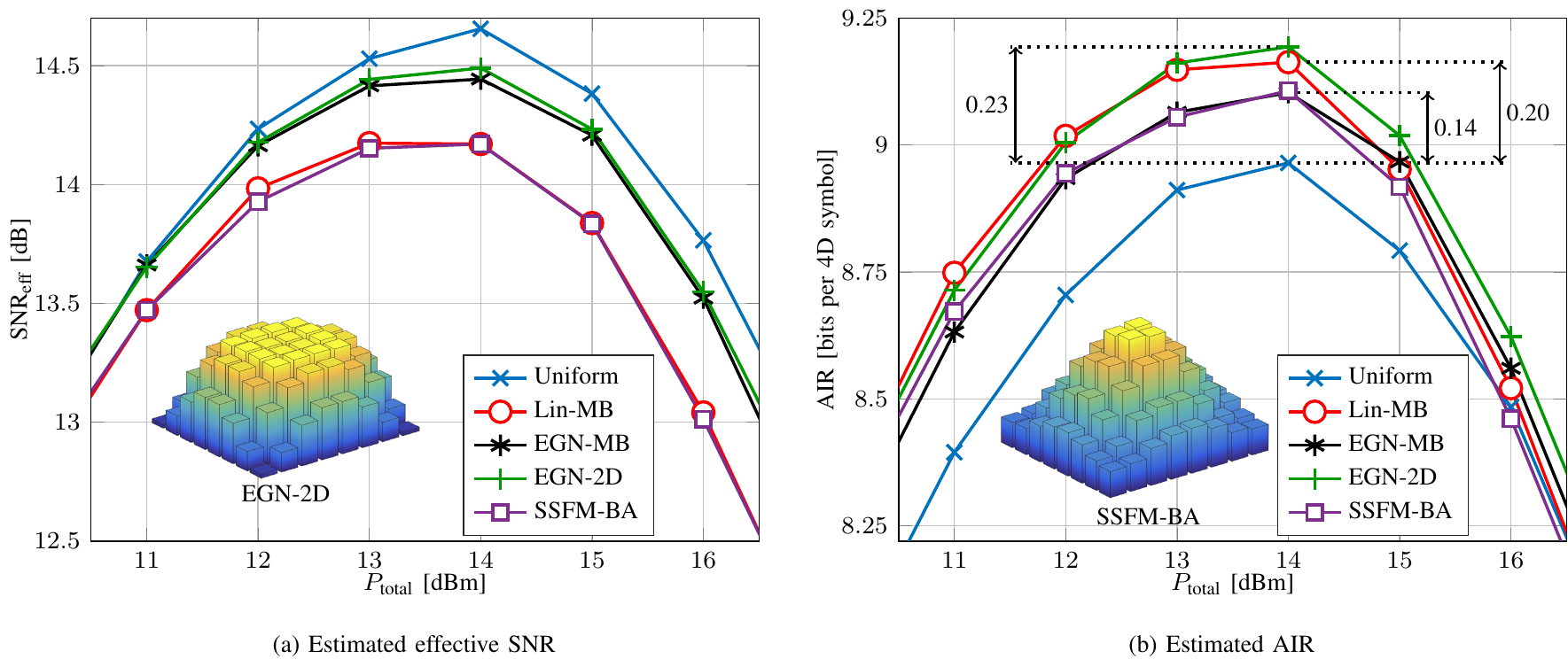}
\caption{Experimental result of the $64$QAM transmission with 9 WDM channels over $180$ km. Inset: \newUn{EGN-2D and SSFM-BA shaped input distribution. The irregular shape of the SSFM-BA optimized PMF is due to finite sequence length during optimization (see Tables~\ref{tab:system_parameters_sim_for_exp1_64} and \ref{tab:shaping-methods}).}}
\label{fig:9WDM_64QAM}
\end{figure*}
\begin{table}
        \caption{\newUn{Properties} of $64$QAM input distributions shaped for $9$ WDM channels transmission}
        \begin{center}
                \setlength{\tabcolsep}{4pt} 
        \renewcommand{\arraystretch}{1.2}
                \begin{tabular}{ r | c | c | c | c | c}
                                                                        & Uniform &   EGN-2D & EGN-MB  & SSFM-BA & Lin-MB \\
                        \hline
                        \hline
                        $\hat{\mu}_4$ & $1.3810$  & $1.4306$ & $1.4705$ & $1.5989$ & $1.6158$ \\
                        \hline
                        $\hat{\mu}_6$ & $2.2258$   & $2.4285$ & $2.5916$ & $3.1782$ & $3.2676$ \\
                        \hline
                        \newUn{$H(X)$} & \newUn{$6.0000$} & \newUn{$5.8191$} & \newUn{$5.9561$} & \newUn{$5.8501$} & \newUn{$5.7529$} \\
                        \hline
                        \newUn{$\delta_{\text{SNR}}$} & \newUn{$0.80$} & \newUn{$0.27$} & \newUn{$0.41$} & \newUn{$0.18$} & \newUn{$0.07$}
                \end{tabular}
        \end{center}
        \label{tab:moments_64_9wdm}
\end{table}

For transmission over $180$ km SMF, the received effective SNR as a function of the total launch power $P_{\text{total}}$ is depicted in Fig.~\ref{fig:9WDM_64QAM}(a). For a small launch power (outside the range shown in Fig.~\ref{fig:9WDM_64QAM}(a)), the effective SNR of all methods is approximately the same. By increasing $P_{\text{total}}$, we observe that the impact of nonlinearity increases, resulting in effective SNR differences depending on the used input. At the optimal launch power $P_{\text{opt}}$ (since the distributions are kept constant for each power sweep, the SNR-maximizing $P_{\text{total}}$ is equal to AIR-maximizing $P_{\text{total}}$) and in the nonlinear regime, the uniform input has higher effective SNRs than the shaped inputs. The EGN-optimized PMFs are penalized by an effective SNR loss of $0.25$~dB and Lin-MB and SSFM-BA are penalized by $0.5$~dB at $P_{\text{opt}}$. The effective SNR losses are attributed to stronger NLI due to an increased peak-to-average power ratio of the transmit signal. By comparing $\hat{\mu}_4$ and $\hat{\mu}_6$ of the used distributions, see Table~\ref{tab:moments_64_9wdm}, we can see that the shaped PMFs with higher moments show lower effective SNR. This experiment confirms \eqref{eq:NLi_noise} and the results of \cite{Fehenberger2016} in the point that increasing $\hat{\mu}_4$ and $\hat{\mu}_6$ results in an SNR penalty.

In Fig.~\ref{fig:9WDM_64QAM}(b), 
the AIR is shown as a function of $P_{\text{total}}$. In the linear regime, all methods perform similarly. At the optimal launch power, the EGN-2D shaping achieves the highest information gain followed by Lin-MB shaping, SSFM-BA shaping and EGN-MB shaping. By considering Fig.~\ref{fig:9WDM_64QAM}(a) and Fig.~\ref{fig:9WDM_64QAM}(b) jointly, 
it is observed that the used shaping methods decrease the effective SNR but increase the AIR compared to uniform. We observe that all shaping methods achieve an information gain, with EGN-2D shaping performing best. \newUn{The relative additional shaping gain by EGN-2D over Lin-MB is $(0.23 - 0.2)/0.2 = 15\%$.}

\begin{figure}
\centering
\includegraphics{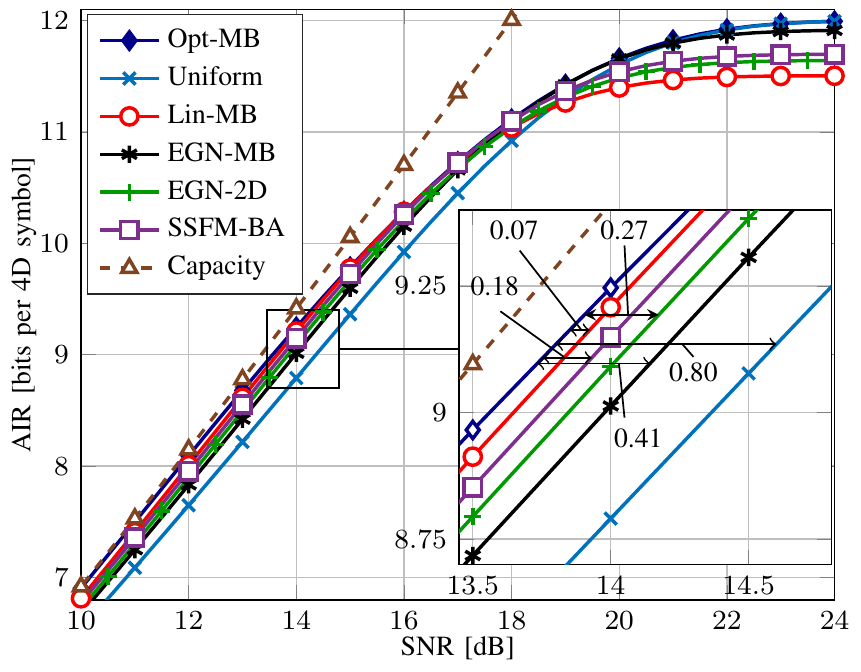}
\caption{\newUn{AIRs for the AWGN channel using the distributions of the 64QAM and 9 WDM channel experiment. Inset: Zoom into the operating region.}}
\label{fig:AWGN_9WDM_64QAM}
\end{figure}

\newUn{AWGN simulations with the tested distributions are performed to investigate whether any transmission improvements are solely the result of shaping for the AWGN channel or due to a trade-off between shaping gain and effective SNR of the fiber channel. The results are shown in Fig.~\ref{fig:AWGN_9WDM_64QAM}, where Opt-MB are MB distributions that are optimized for each SNR. Since the PMF obtained by Opt-MB is known to nearly achieve the capacity of the AWGN channel with constrained-size constellations}\new{\cite{Wu2012_optMB}}\newUn{, it is used to determine the shaping gap $\delta_{\text{SNR}}$ for all other considered distributions (which are optimized for one particular SNR). The shaping gaps and the entropies of the input distributions $H(X)$ are given in Table~\ref{tab:moments_64_9wdm}. The simulation results show that the AWGN shaping gaps of the applied distributions vary strongly, with Lin-MB being closest to Opt-MB, cf. inset of Fig.~\ref{fig:AWGN_9WDM_64QAM} and Table~\ref{tab:moments_64_9wdm}. We conclude that any improvement by EGN-2D shaping over Lin-MB results from an improved trade-off between a shaping gain for the AWGN channel and a reduced effective SNR loss for the nonlinear fiber channel.}

\subsection{256QAM Transmission with 9 WDM Channels}

\begin{figure*}
\centering
\includegraphics{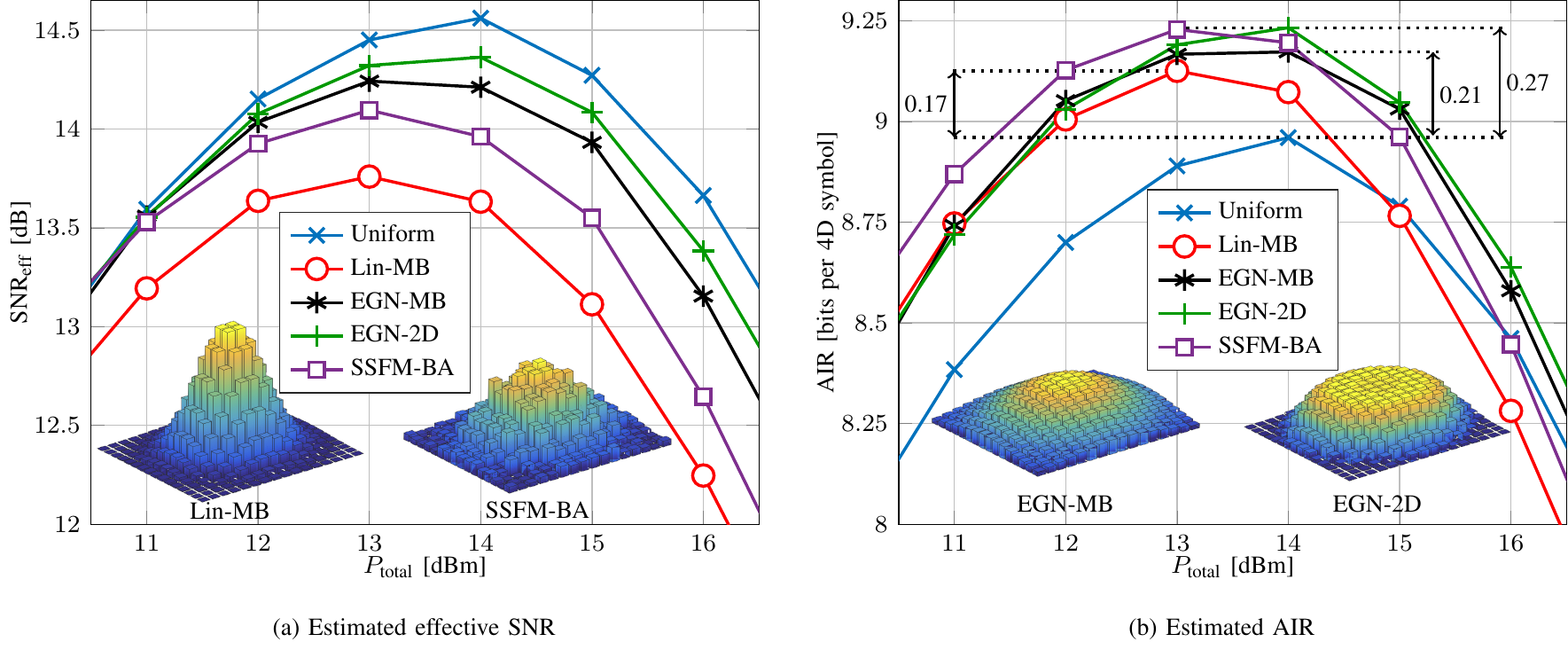}
\caption{Experimental result of the $256$QAM transmission with 9 WDM channels over $180$ km. Insets: shaped input distributions. \newUn{The irregular shape of the SSFM-BA optimized PMF is due to finite sequence length during optimization (see Tables~\ref{tab:system_parameters_sim_for_exp1_64} and \ref{tab:shaping-methods}).}}
\label{fig:9WDM_256QAM}
\end{figure*}

\begin{table}
        \caption{\newUn{Properties} of $256$QAM input distributions shaped for $9$ WDM channels transmission}
        \begin{center}
                \setlength{\tabcolsep}{4pt} 
        \renewcommand{\arraystretch}{1.2}
                \begin{tabular}{ r | c | c | c | c | c}
                                                                        & Uniform &   EGN-2D & EGN-MB  & SSFM-BA & Lin-MB \\
                        \hline
                        \hline
                        $\hat{\mu}_4$ & $1.3953$  & $1.4705$ & $1.5915$ & $1.8125$ & $1.9191$ \\
                        \hline
                        $\hat{\mu}_6$ & $2.2922$   & $2.6350$ & $3.1612$ & $4.4049$ & $5.2034$ \\
                        \hline
                        \newUn{$H(X)$} & \newUn{$8.0000$} & \newUn{$7.4831$} & \newUn{$7.8095$} & \newUn{$7.5146$} & \newUn{$6.8595$} \\
                        \hline
                        \newUn{$\delta_{\text{SNR}}$} & \newUn{$0.88$} & \newUn{$0.32$} & \newUn{$0.28$} & \newUn{$0.11$} & \newUn{$0.004$}
                \end{tabular}
        \end{center}
        \label{tab:moments_256_9wdm}
\end{table}

For the following analysis, a 256QAM constellation is considered with the same signal and system parameters as above. The received effective SNR as a function of $P_{\text{total}}$ is shown in Fig.~\ref{fig:9WDM_256QAM}(a). 
We observe that methods with higher moments show lower effective SNR, cf. Table \ref{tab:moments_256_9wdm}, which verifies also for 256QAM that increasing $\hat{\mu}_4$ and $\hat{\mu}_6$ results in an effective SNR penalty.
The AIR as a function of $P_{\text{total}}$ is depicted in Fig.~\ref{fig:9WDM_256QAM}(b). 
In the linear regime, Lin-MB, EGN-MB and EGN-2D perform equally well and SSFM-BA achieves a higher gain. At the optimal launch power, EGN-2D shaping and SSFM-BA shaping achieve the highest information gain of $0.27$~bits per 4D symbol, followed by EGN-MB shaping and Lin-MB shaping with $0.21$~bits per 4D symbol and $0.17$~bits per 4D symbol, respectively. \newUn{The relative shaping gain by EGN-2D over Lin-MB is $59\%$.}

\newUn{The performance gain of SSFM-BA over Lin-MB in the linear regime results from two facts. Firstly, as shown in Fig.~\ref{fig:9WDM_256QAM}(a), the effective SNR of Lin-MB is smaller than for the other distributions due to increased NLI. Secondly, as pointed out also in}\new{\cite{Yankov16_exp}}\newUn{, Lin-MB is more sensitive to a deviation from the optimal launch power than a BA-optimized distribution.}

The similar information gain of SSFM-BA shaping and EGN-2D shaping at the optimal launch power but their different performances in the linear and nonlinear regime can be explained as follows.
While SSFM-BA shaping achieves a large information gain in the linear regime by a distribution that has a distinct peak around the origin, cf. inset in Fig.~\ref{fig:9WDM_256QAM}(a) \newUn{and $\delta_{\text{SNR}}$ in Table \ref{tab:moments_256_9wdm}}, it is penalized in the nonlinear regime for its increased $\hat{\mu}_4$ and $\hat{\mu}_6$. In contrast, EGN-2D shaping has a smaller gain in the linear regime due to its \newUn{inverse bathtub shape}, cf. inset in Fig.~\ref{fig:9WDM_256QAM}(b) \newUn{and Table \ref{tab:moments_256_9wdm}}, but performs better in the nonlinear regime as a result of its smaller moments. In combination, these features result in a similar information gain of both approaches at the optimal launch power.

The optimal launch power is reduced by Lin-MB shaping and SSFM-BA shaping, whereas for EGN-MB shaping and EGN-2D shaping the optimal launch power is the same as for the uniform input. This effect can be explained by the EGN model \eqref{eq:NLi_noise} and the order of the shaping methods with respect to their $\hat{\mu}_4$ and $\hat{\mu}_6$. The parameters $\chi_0$, $\chi_4$, $\chi_4'$ and $\chi_6$ are independent of $P_{\text{tx}}$ and $P_X$, which means the higher $\hat{\mu}_4$ and $\hat{\mu}_6$ the smaller $P_{\text{opt}}$, see (\ref{eq:effective_SNR}) and (\ref{eq:NLi_noise}). Thus, the order of the shaping methods with respect to $P_{\text{opt}}$ is reverse to their order with respect to $\hat{\mu}_4$ and $\hat{\mu}_6$. We conclude that by taking the nonlinear effects for shaping into account, i.e., EGN-MB shaping, EGN-2D shaping and SSFM-BA shaping, we can slightly increase the information gain over Lin-MB. Furthermore, SSFM-BA shaping and EGN-2D shaping perform similarly.


\subsection{Fully Loaded Optical Transmission System}
Since fully loaded systems are a highly relevant configuration for practical fiber communication, probabilistic shaping for a C-band transmission with $80$ WDM channels over $160$ km is investigated in the following. Simulations of fully loaded systems based on SSFM models are very time consuming due to the large simulation bandwidth. Therefore, SSFM-BA optimization is not practical with standard computers and could not be included in this comparison. However, a simulation based on the EGN model can be done in an acceptable time which means that EGN-MB shaping as well as EGN-2D probabilistic optimization are possible.
\begin{table}
        \caption{Signal parameters of the fully loaded transmission system}
        \begin{center}
                \begin{tabular}{ r | l }
                        Modulation & $256$QAM \\
                        Symbol rate $R_{X}$ & $20$ GBd \\
                        Pulse shaping & RRC filter \\
                        RRC roll-off $\rho$ & $0.5$ \\
                        WDM channel number $N_{\text{WDM}}$ & $80$ \\
                        WDM channel spacing $f_{\text{WDM}}$ & $50$ GHz \\
                        Polarization & dual \\
                        \newUn{Block length} & \newUn{$13200$ symbols} \\
                        \newUn{Number of blocks} & \newUn{$40$ -- $120$ } \\
                        Pilot rate & $10\%$ \\
                \end{tabular}
        \end{center}
        \label{tab:signal_parameters_exp2}
\end{table}
\begin{figure}
        \begin{center}
                \includegraphics{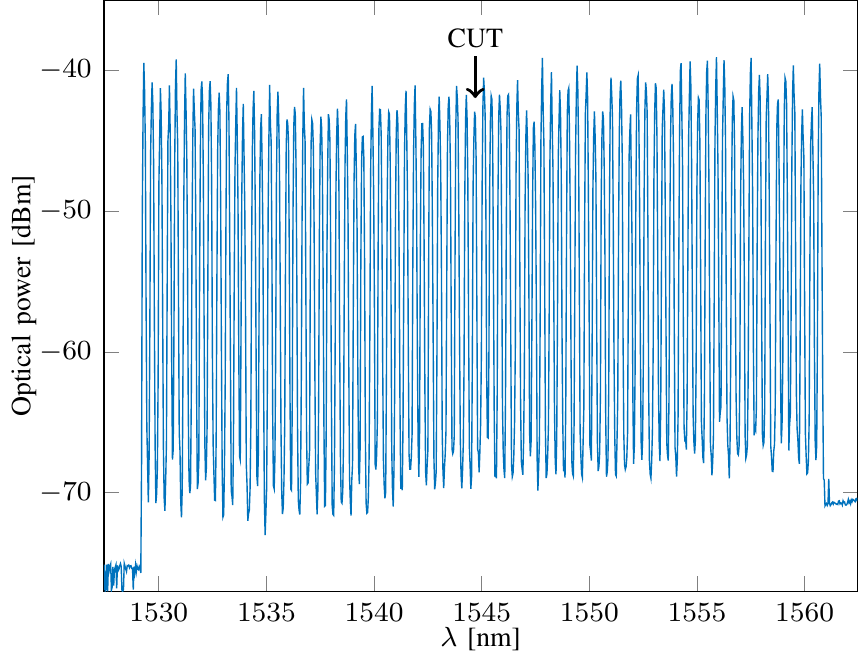}
        \end{center}   
        \vspace{-0.5cm}   
        \caption{Spectrum of the $80$ WDM channels transmitted over $160$ km SMF.}
        \label{Exp:ch_80_spec}
\end{figure}


\begin{figure*}
\centering
\includegraphics{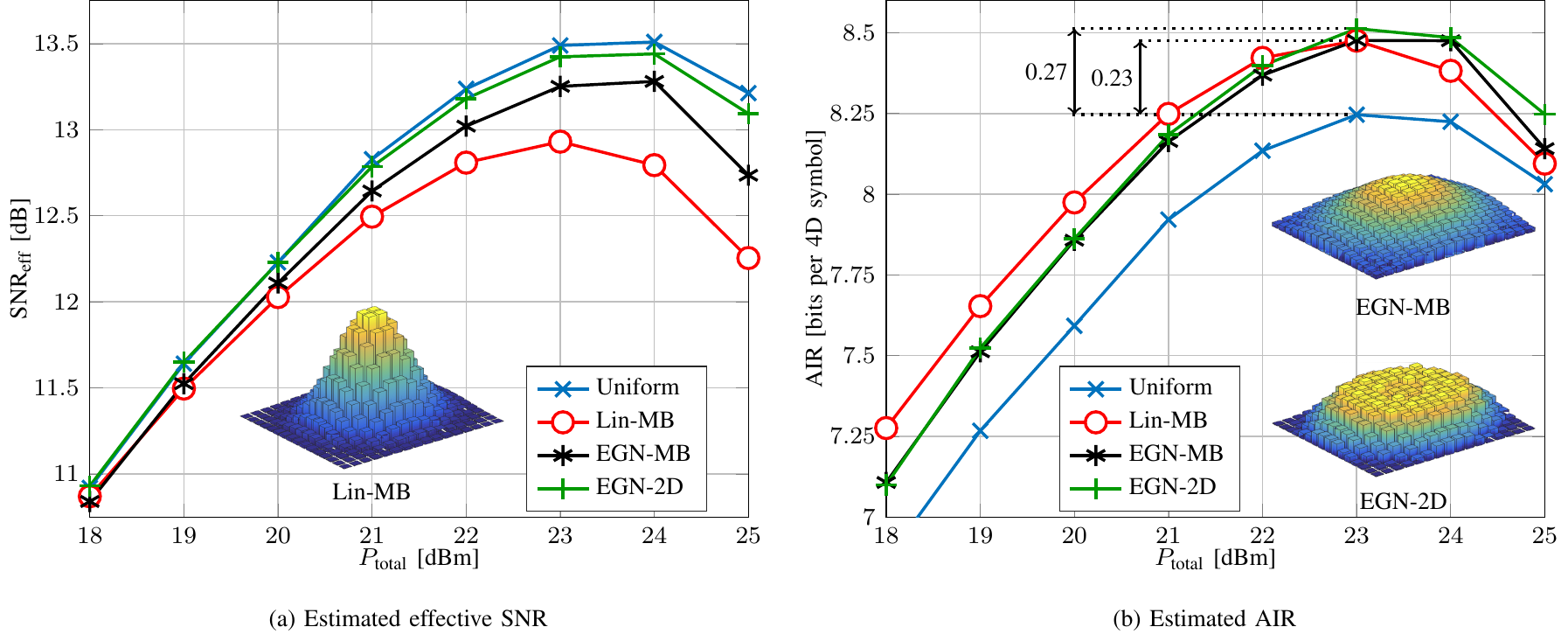}
\caption{Experimental result of the $256$QAM transmission with 80 WDM channels over $160$ km. Insets: shaped input distributions.}
\label{fig:80WDM_256QAM}
\end{figure*}

\begin{table}
        \caption{\newUn{Properties} of $256$QAM input distributions shaped for $80$ WDM channels transmission}
        \begin{center}
                \setlength{\tabcolsep}{4pt} 
        \renewcommand{\arraystretch}{1.2}
                \begin{tabular}{ r | c | c | c | c }
                                                                        & Uniform &   EGN-2D & EGN-MB & Lin-MB \\
                        \hline
                        \hline
                        $\hat{\mu}_4$ & $1.3953$  & $1.4536$ & $1.6003$  & $1.8819$ \\
                        \hline
                        $\hat{\mu}_6$ & $2.2922$   & $2.5615$ & $3.2047$ & $4.9105$ \\
                        \hline
                        \newUn{$H(X)$} & \newUn{$8.0000$} & \newUn{$7.6496$} & \newUn{$7.7943$} & \newUn{$7.0339 $} \\
                        \hline
                        \newUn{$\delta_{\text{SNR}}$} & \newUn{$0.79$} & \newUn{$0.35$} & \newUn{$0.25$} & \newUn{$0.006$}
                \end{tabular}
        \end{center}
        \label{tab:moments_256_80wdm}
\end{table}

Table \ref{tab:system_parameters_sim_for_exp1_64} and Table \ref{tab:signal_parameters_exp2} show the system parameters and the signal parameters used for the optimization of the input distribution, respectively. \newUn{The experimental setup is very similar to the one described in Sec.~\ref{section3}, and the only differences to the preceding Secs.~\ref{section4}-A to \ref{section4}-C are that the total number of channels is $80$, the WDM grid is changed to $50$ GHz and the symbol rate is increased to $20$ GBd in order to achieve similar bandwidth occupancy and spectral efficiency as in the 9 WDM channel case. The 80 channels were equalized according to a trade-off between power- and OSNR equalization resulting in all the interferers to be within $\pm 2.5$ dB of the channel under test.} The most important parameters of the experimental setup are listed in Table \ref{tab:signal_parameters_exp2}, and the spectrum of the transmitted WDM signal is plotted in Fig.~\ref{Exp:ch_80_spec}. The received SNR of the different shaping strategies as a function of $P_{\text{total}}$ is shown in Fig.~\ref{fig:80WDM_256QAM}(a). 
As in the previous experiments with 9 WDM channels, increasing $\hat{\mu}_4$ and $\hat{\mu}_6$, cf. Table \ref{tab:moments_256_80wdm}, results in an effective SNR penalty also for full C-band transmission.
The estimated AIR as a function of $P_{\text{total}}$ is depicted in Fig.~\ref{fig:80WDM_256QAM}(b). 
In the linear regime, all shaping methods achieve an information gain, with Lin-MB shaping performing best. The EGN-based shapings perform similarly but slightly worse compared to Lin-MB shaping. At the optimal launch power, EGN-2D shaping achieves the highest information gain of $0.27$~bits per 4D symbol, followed by EGN-MB shaping and Lin-MB shaping with a gain of $0.23$~bits per 4D symbol. Since $P_{\text{total}}$ was limited to $25$~dBm, we could not evaluate the performance of the methods in the highly nonlinear regime. We observe that a similar information gain at $P_{\text{opt}}$ is achieved for a fully loaded system by Lin-MB shaping and the EGN-based shaping methods. A potential reason for this could be that the employed EGN-based PMF optimization slightly overestimates the NLI variance in comparison to the experimental setup, which makes the resulting distributions tailored for a nonlinear regime that is not experienced in the experiments. \newUn{At $P_{\text{opt}}$, we observe a relative additional gain of $17\%$ by EGN-2D over Lin-MB.}



\subsection{Future work}
The nonlinear fiber channel induces strong temporal correlations of the signal by the interplay of dispersive memory and inter-channel nonlinear crosstalk \cite{Dar2013OptExp_PropertiesNLIN,Fehenberger_end}. This implies that the used methods are suboptimal as the considered symbol-by-symbol shaping does not exploit this memory. In \cite{Geller2016} and \cite{Yankov2017_TempShaping}, schemes that shape symbols transmitted over the same WDM channel at consecutive time slots jointly in order to take the memory of the channel into account are presented. The optimization in \cite{Yankov2017_TempShaping} requires SSFM simulations, which are time consuming. Adopting the EGN optimization strategy described in Sec.~\ref{sec:shaped_dists} to temporal statistics is of interest for future research.

\section{Conclusion}
In this paper, various methods to find optimized probabilistically shaped QAM distributions have been studied in a back-to-back setup and in unrepeated transmission experiments. It was shown that the effective SNR of the back-to-back configuration was not affected by the input PMF, indicating that with the employed modulation format independent DSP, shaping does not cause an implementation penalty. This allows us to conclude that any SNR difference in the optical transmission experiments is caused by nonlinear fiber effects. Different shaping methods are compared for unrepeated transmission systems with $64$QAM and $256$QAM and different WDM channel count ($9$ channels and fully-loaded system). Increasing the fourth and sixth moment of the input distribution is demonstrated to decrease the effective SNR in all cases, which experimentally confirms the EGN model in this regard. \newUn{Although the effective SNR is decreased for the variety of examined input PMFs, all shaping methods achieve relatively small absolute AIR improvements. However, the relative shaping gain by EGN-2D over Lin-MB is found to be significant, amounting to up to $59\%$. Since the optimization complexity of the EGN-based methods is comparable to Lin-MB, shaping with EGN-optimized input distributions should be used for short-reach systems when the resources for shaping are available.}

\section*{Acknowledgments}
The authors would like to thank Dr. Alex Alvarado (TU/e) for his comments that greatly helped to improve the manuscript.



\end{document}